\documentclass{jfm}

\usepackage{CJK}

\usepackage{ulem}
\usepackage[T1]{fontenc}
\usepackage{calligra}

\usepackage{graphicx}
\usepackage{dcolumn}
\usepackage{bm}

\usepackage{times}
\usepackage{verbatim}
\usepackage{graphicx}
\usepackage{graphics,epsfig}

\usepackage{theorem}
\usepackage{makeidx}
\usepackage{amsmath}
\usepackage{epic}
\usepackage{amscd}

\usepackage{bbm}
\usepackage{xy}
\usepackage{amssymb}
\usepackage{epsfig}

\usepackage{cancel}

\usepackage{graphicx}
\usepackage{natbib}

\ifCUPmtlplainloaded \else
  \checkfont{eurm10}
  \iffontfound
    \IfFileExists{upmath.sty}
      {\typeout{^^JFound AMS Euler Roman fonts on the system,
                   using the 'upmath' package.^^J}%
       \usepackage{upmath}}
      {\typeout{^^JFound AMS Euler Roman fonts on the system, but you
                   dont seem to have the}%
       \typeout{'upmath' package installed. JFM.cls can take advantage
                 of these fonts,^^Jif you use 'upmath' package.^^J}%
      }
  \else
  \fi
\fi

\ifCUPmtlplainloaded \else
  \checkfont{msam10}
  \iffontfound
    \IfFileExists{amssymb.sty}
      {\typeout{^^JFound AMS Symbol fonts on the system, using the
                'amssymb' package.^^J}%
       \usepackage{amssymb}%

      }{}
  \fi
\fi

\ifCUPmtlplainloaded \else
  \IfFileExists{amsbsy.sty}
    {\typeout{^^JFound the 'amsbsy' package on the system, using it.^^J}%
     \usepackage{amsbsy}}
    {}
\fi

\newsavebox{\astrutbox}
\sbox{\astrutbox}{\rule[-5pt]{0pt}{20pt}}

\title[
Isotropic polarization of compressible flows
]{
Isotropic polarization of compressible flows
}

\author[J.-Z. Zhu
]%
{Jian-Zhou Zhu}

\affiliation{Su-Cheng Centre for Fundamental and Interdisciplinary Sciences, Gaochun, Nanjing 211316 China and Life and Chinese Medicine Study Center, Gui Lin Tang Lab., Yong'an, Fujian 366025 China
}

\pubyear{2010}
\volume{650}
\pagerange{119--126}
\date{?; revised ?; accepted ?. - To be entered by editorial office}
\begin{document}

\maketitle

\begin{abstract}
The helical absolute equilibrium of a compressible adiabatic flow presents not only the polarization between the two purely helical modes of opposite chiralities but also that between the vortical and acoustic modes, deviating from the equipartition predicted by {\sc Kraichnan, R. H.} [1955
The Journal of the Acoustical Society of America {\bf 27}, 438--441.].
Due to the existence of the acoustic mode, even if all Fourier modes of one chiral sector in the sharpened Helmholtz decomposition [{\sc Moses, H. E.} 1971
SIAM ~(Soc. Ind. Appl. Math.) J. Appl. Math. {\bf 21}, 114--130] are thoroughly truncated, leaving the system with positive definite helicity and energy, negative temperature and the corresponding large-scale concentration of vortical modes are not allowed, unlike the incompressible case.
\end{abstract}

\section{Introduction}
``Local anisotropy' or ``polarization'', referring to the dependence on the directions in the local coordinate frame of the wave vector space, is one of the important characteristics of turbulence \citep[][and references therein]{CambonJacquinJFM89}. Such polarization can be due to either linear (damping or wave) or nonlinear mechanisms (helicity, say), or both, as appears in rotating flows and quasi-static magnetohydrodynamics, among others \citep{CambonGodeferd93}.
\citet[][`K55' hereafter]{k55} studied the absolute equilibrium (AE) of an adiabatic compressible flow model and showed with small-excitation approximation the complete equipartition of energy among all the vortical, dilatational and density modes [similar result for the incompressible case had been obtained by \citet{Lee1952} and \citet{Hopf52}, the latter of which applied functional calculus to formally derive it, but without explicitly introducing the necessary Galerkin truncation], and non-equipartition was anticipated for the strong-excitation case.
\citet{KrstulovicETC09} proposed a new algorithm to generate the irrotational, thus nonhelical, compressible AE, and they indeed observed nearly Gaussian and equipartition at lower temperatures but non-Gaussian statistics and non-equipartitions of total compressible energies, together with ``non-physical'' negative fluid density for higher temperatures. K55 also discussed the effects of polarized dissipation. Note that helicity is (formally) conserved in barotropic flows \citep{Moffatt69}, thus, according to \citet{k73}, incompleteness of K55's analysis. Also note that \citet{k73} refuted the speculation of the possibility of inverse cascade in three-dimensional (3D) helical turbulence from the superficial analogy with 2D flows bearing also two ideal invariants, which may lead to the belief that two positive-definite invariants must indicate inverse cascade: We will update K55 by additional respect of helicity and show that the acoustic mode in the compressible flow excludes such a possibility. We should notice that such multiple-constraint analysis based on time-reversible system could have relevant indications for nonequilibrium and irreversible ensembles \citep{nonir} of other types of time-reversible modified Navier-Stokes equations \citep{SheJackson} as we will come back in the end.

Let $\bm{u}$, $\rho$, $\eta$ and $c$ represent the flow velocity, density, viscosity and velocity of sound in a compressible flow with the pressure $p$ given by the adiabatic relation $p=c^2\rho$. Since now, $p$ is independent of the temperature (otherwise the temperature should be fixed, i.e., iso-temperature), the mass and momentum conservation laws are decoupled from the energy equation, forming an autonomous system
where the entropy modes are not concerned \citep{SagautCambon08}. 
K55 introduced the transformation $\rho=\rho_0\exp\{\zeta\}$ (`Kraichnan transformation' hereafter) with $\rho_0$ being the equilibrium density, which, in the forceless case, leads to the equations of motion
\begin{eqnarray}
  \dot{\zeta}+\zeta_{,\alpha}u_{\alpha}+u_{\alpha,\alpha} &=& 0 \label{eq:k1}\\
  \dot{u}_{\lambda} + u_{\sigma}u_{\lambda,\sigma}+c^2\zeta_{,\lambda}-\eta \theta_{\lambda\sigma,\sigma}&=& 0, \label{eq:k2}
\end{eqnarray}
where $\theta_{\alpha \beta}=u_{\alpha,\beta}+u_{\beta,\alpha}-\frac{2}{3}\delta^{\alpha}_{\beta}u_{\sigma,\sigma}$, $(\bullet)_{,\gamma}=\partial (\bullet)/\partial x^{\gamma}$ and $\dot{(\bullet)}=\partial_t (\bullet)$.
Working in a cyclic box of dimension $2\pi$ with $V=[0,2\pi)^3$ and applying Fourier representation for all the dynamical variables $v(\bm{r}) \to \hat{v}(\bm{k})$, say, $\bm{u}(\bm{r})=\sum_{\bm{k}}\hat{\bm{u}}(\bm{k}) \exp\{\hat{i}\bm{k}\cdot \bm{r} \}$ with $\hat{i}^2=-1$, K55 constructed a phase space by the real and imaginary parts of $v$s and showed that the flow in such a phase space is incompressible, the Liouville theorem, in the inviscid case.
K55 further took the `small-excitation' approximation for the ideal invariant of the total, kinetic plus potential, mean energy per unit mass: 
\begin{equation}\label{eq:E}
E=\frac{1}{2} \sum_{\bm{k}}[\hat{u}_{\lambda}(\bm{k})\hat{u}^*_{\lambda}(\bm{k})+c^2|\hat{\zeta}(\bm{k})|^2]
\end{equation}
which, according to Eqs. (\ref{eq:k1}) and (\ref{eq:k2}) with the viscous terms removed, obeys in $\bm{r}$-space $\dot{E}+\langle u_{\alpha}e_{,\alpha}\rangle+c^2\langle(\zeta u_{\alpha})_{,\alpha}\rangle=0$ with $e=u_{\sigma}u_{\sigma}+c^2\zeta^2$ and $\langle \bullet \rangle=\frac{1}{(2\pi)^3}\int_V \bullet dV$. The periodic boundary condition kills $\langle(\zeta u_{\alpha})_{,\alpha}\rangle$, and $E$ is not rigorously an invariant but Kraichnan's approximation of the total energy. Galerkin truncation, say, imposing all modes with $|\bm{k}|$ greater than some cut-off value $K$ to be zero, is then performed, which does not change the approximated conservation of energy due to its being quadratic and diagonal in $\bm{k}$ \citep{k73}. Reasoning with the $H$-theorem, K55 predicted that an ensemble of systems will tend towards an equilibrium distribution uniform over the surfaces of constant energy, thus an equipartition, in particular between the shear/vortical and acoustic modes: Such K55 fields are isotropic, both globally/`directionally' (independent on the direction of $\bm{k}$) and `locally' (independent on the directions in the local coordinate frame), in the terminology of \citet{CambonJacquinJFM89}.
\citet{Moffatt69} has shown that, with the vorticity defined by $\bm{\omega}(\bm{r})=\nabla\times\bm{u}(\bm{r})=\sum_{\bm{k}}\hat{\bm{\omega}}(\bm{k})\exp\{\hat{i}\bm{k}\cdot\bm{r}\}$, the helicity 
\begin{equation}\label{eq:H}
H=\frac{1}{2} \sum_{\bm{k}}\hat{\bm{u}}(\bm{k}) \cdot \hat{\bm{\omega}}(-\bm{k})
\end{equation}
is also formally an invariant in barotropic flows with $p=p(\rho)$ and, just as in the incompressible case \citep{k73}, should be respected. It is also desirable to extend the polarization analysis to cases with strong excitations \citep{ShivamoggiEPL97,KrstulovicETC09}, with \textit{ad hoc} assumption or uncontrollable approximation, but we will only focus on the polarization issue, using still the small-excitation approximation for preciseness and analytical tractability. Our purpose is to deliver a crystal clear message of the so-called `helicity effect' in such a situation, instead of some heroic theory or the final solution of the whole problem (with general equation of state and strong excitations) which appears extremely complex, as we will further remark later: indeed, since the absolute equilibrium is nevertheless far from the turbulent state (except for the large scales in the inverse cascade case) and in general is for qualitative theoretical insights, it would hardly be more illuminating for our purpose to further write down or compute the very complicated strong-excitation/high-temperature absolute equilibria.

\section{Compressible helical absolute-equilibrium polarization and non-equipartition: differences with the incompressible case}

\citet{Moses71} has `sharpened' the Helmholtz theorem by further decomposing the transverse (`solenoidal'/`vortical') velocity field into a left-handed and right-handed chiral modes of sign-definite helixity (helicity intensity)
\begin{equation}\label{eq:hd}
\hat{\bm{u}}(\bm{k})=\hat{u}_+(\bm{k})\hat{\bm{h}}_+(\bm{k})+\hat{u}_-(\bm{k})\hat{\bm{h}}_-(\bm{k})+\hat{u}_|(\bm{k})\bm{k}/k,
\end{equation}
with $\hat{i}\bm{k}\times \hat{\bm{h}}_s=sk\hat{\bm{h}}_s$ and $s=\pm$ (denoting opposite --- right- v.s. left-handed --- screwing directions, or \textit{chiralities}, around $\bm{k}$), and that
\begin{equation}\label{eq:ce}
E=\frac{1}{2}\sum_{\bm{k}}|\hat{u}_+|^2+|\hat{u}_-|^2+|\hat{u}_||^2+c^2|\hat{\zeta}|^2,
\end{equation}
for small excitation as in K55, and
\begin{equation}\label{eq:ch}
H=\frac{1}{2}\sum_{\bm{k}} k|\hat{u}_+|^2-k|\hat{u}_-|^2.
\end{equation}
The equations for incompressible flows (without the longitudinal/dilational components) corresponding to Eqs. (\ref{eq:hd},\ref{eq:ce},\ref{eq:ch}) were used by \citet{CambonJacquinJFM89} and \citet{W92} to study rotating turbulence and triad instabilities etc. Denoting $\hat{\bm{u}}_{\perp}(\bm{k})=\hat{u}_+(\bm{k})\hat{\bm{h}}_+(\bm{k})+\hat{u}_-(\bm{k})\hat{\bm{h}}_-(\bm{k})$, we can rewrite Eqs. (\ref{eq:k1}) and (\ref{eq:k2}) without viscosity in the following way
\begin{eqnarray}
  \dot{\hat{\zeta}}(\bm{k}) &=& \sum_{\bm{p}+\bm{q}=\bm{k}} [\hat{\zeta}(\bm{p})\bm{p}\cdot\hat{\bm{u}}_{\perp}(\bm{q})+ \hat{\zeta}(\bm{p})\bm{p}\cdot\hat{\bm{u}}_{|}(\bm{q})] + k\hat{u}_|(\bm{k}) \label{eq:z1}\\
  \dot{\hat{\bm{u}}}_|(\bm{k}) &=& \hat{\bm{N}}_|(\bm{k}) + c^2\bm{k}\hat{\zeta}(\bm{k}) \label{eq:z2}\\
  \dot{\hat{\bm{u}}}_{\perp}(\bm{k}) &=& \hat{\bm{N}}_{\perp}(\bm{k}) \label{eq:z3},
\end{eqnarray}
with $\bm{N}=\bm{u}\cdot\nabla\bm{u}=(\bm{u}_|+\bm{u}_{\perp})\cdot\nabla(\bm{u}_|+\bm{u}_{\perp})$, showing the complicated interactions [with, say, $\hat{\bm{N}}_|(\bm{k})=\widehat{(\bm{u}_{\perp}\cdot\nabla\bm{u}_{\perp})}_|+\widehat{(\bm{u}_{|}\cdot\nabla\bm{u}_{|})}_|+\widehat{(\bm{u}_{\perp}\cdot\nabla\bm{u}_{|})}_|+\widehat{(\bm{u}_{|}\cdot\nabla\bm{u}_{\perp})}_|$], of the acoustic and vorticity modes.
Note that the terms $\hat{\zeta}(\bm{p})\bm{p}\cdot\hat{\bm{u}}_{\perp}(\bm{q})$ and $\widehat{(\bm{u}_{\perp}\cdot\nabla\bm{u}_{\perp})}_{\perp}$ as in the incompressible case conserve, for each interacting triad, $\langle\zeta^2\rangle$ and $\langle \bm{u}_{\perp}^2\rangle$ respectively. Small-excitation approximation for the total energy in terms of $\zeta$ and $\bm{u}$ is indeed necessary to proceed with the analytical calculation. \citet{Moses71} solved the corresponding linearized problem, and K55's treatment can be viewed as quasi-linear [with $c^2=p'(\rho_0)$ being the constant of the background field $\rho_0$.]
The inviscid and forceless Galerkin truncated dynamics for absolute equilibrium analysis may also be viewed with the balance of the nonvanishing internal viscous force and the external stirring force $\bm{f}(t)$. It is trivial to take the external force be the internal viscous force with constant viscosity $\eta$, but it may be more interesting to modify the viscous term to balance the independent external forcing: One particular possibility is, say, replacing the $\eta$ neglected in Eq. (\ref{eq:z2}) for the longitudinal component with a dynamical eddy viscosity $\eta_|({\bm{k}},t)$, i.e.,
\begin{equation}\label{eq:de}
    \eta \to \eta_|({\bm{k}})=3\hat{\bm{f}}_|({\bm{k}})\cdot \hat{\bm{u}}_|^{*}({\bm{k}})/[4k^2 |\hat{\bm{u}}_|({\bm{k}})|^2],
\end{equation}
and similarly $\eta_{\pm}$ for the left and right chiral sectors of Eq. ({\ref{eq:z3}) for the transverse components, \textit{mutatis mutandis}.
Following the nowadays a routine procedure \citep{hydrochirality}, we immediately obtain with the constant of motion $C=\alpha E + \beta H$ from the Gibbs distribution $\sim \exp\{-C\}$ the  absolute equilibrium modal spectra
\begin{eqnarray}
  U^+(\bm{k})\triangleq \langle |\hat{u}_+|^2 \rangle &=& \frac{1}{\alpha+\beta k}, \label{eq:u+}\\
  U^-(\bm{k})\triangleq \langle |\hat{u}_-|^2 \rangle &=& \frac{1}{\alpha-\beta k}, \label{eq:u-} \\
  Z(\bm{k})\triangleq \langle c^2|\hat{\zeta}|^2 \rangle &=& \frac{1}{\alpha},  \\
  U^|(\bm{k})\triangleq \langle |\hat{u_|}|^2 \rangle &=& \frac{1}{\alpha}.
\end{eqnarray}
We see that the spectra are directionally isotropic; but, when $\beta$ or $H$ is not zero energies are polarized, except for the equipartition between the dilatational and density/pressure modes; and, the vortical-mode energy is larger than the `acoustic' one
\begin{equation}\label{eq:neq}
U^{\perp}\triangleq U^{+} + U^{-}> U^{\sim} \triangleq Z+U^|,
\end{equation}
which indicates that helical states tend to have higher vorticity-mode energy and reduce `noise'; see more discussions in the next section. [It is interesting, though probably useless in practice, to note that locally in the plane perpendicular to $\bm{k}$, the statistics are still isotropic, because the polarization between $U^+$ and $U^-$ is about the screwing directions around $\bm{k}$, having no preferred direction perpendicular to it.] A physical question is whether and how such polarization and equipartition of the thermalized state persists in the nonequilibrium turbulence, in particular whether the acoustic-mode energy level stays in between the vortical ones of opposite chiralities.

Note that Eqs. (\ref{eq:u+}, \ref{eq:u-}) are respectively for `purely helical absolute equilibrium' of the `chiroids' $\hat{u}_s\hat{\bm{h}}_s(\bm{k})e^{\bm{k}\cdot\bm{r}}+c.c.$ \citep{hydrochirality}. Due to the existence of the acoustic modes ($u_|$ and $\zeta$), even if we thoroughly remove one chirality, say, that denoted by '$-$', of the vortical modes, as in the numerical simulations of \citet{bmt12} and in the mathematical proof used in \citet{BiferaleTiti}, realizability of acoustic modes (positiveness of $U^|$ and $Z$) excludes the possibility of `negative temperature', $\alpha <0$, unlike the incompressible homochiral case \citep{hydrochirality}; so, there is generally no indication, from such absolute equilibrium, of large-scale concentration or inverse transfer of $U$ for the compressible case, except that the large-scale concentration state with the very specific truncation scheme discussed in Sec. 2.2.2 of \citet{hydrochirality} may still be possible. One thus may say that helicity has less pronounced effects for compressible flows. \citet{KrstulovicETC09} instead further implicitly truncated the rotational modes by using the equations of irrotational flow [otherwise, since the vorticity is not invariant after Galerkin truncation of the original Euler equations, it would not make much sense to talk about compressible irrotational absolute equilibrium, even if with an irrotational initial field, as they did\footnote{Erratum or Clarification: Theoretically, such a statement is not correct, since the initially irrotational state will be kept by the flow; but, in physical reality or numerical simulations, unavoidable vortical noise will indeed require extrally imposing the truncation of the vortical modes.}]: In the numerical results without limiting to the small-excitation-approximation formulation, nearly equipartition at low temperature and obvious non-equipartition and non-Gaussian at high temperature were observed; the formulation they used for more general, but still isoentropic and barotropic, flows applies the transformations of Weber and Clebsch to find the variational principle \citep{Mobbs}, which involves several auxiliary potential functions, making the problem structurally complicated and practically inconvenient. It may nevertheless possible to proceed the calculation and computation, though formidable and difficult to be illuminating [even for the very specific irrotational case, the calculations and computations of \citet{KrstulovicETC09} offer no new physical insight except for the general notion of nonequipartition/non-Gaussianity already expected by K55]. Actually, based on the above fundamental results, various other approaches can be further applied for different specific problems: For example, the decimation schemes by \citet{FrischETCprl12} and the variety of chiral (sub)ensembles in the fashion of \citet{ZhuPoF14} can also be constructed for particular problems.

The effect of the Galerkin truncation on a single solution can be well understood in the one-dimensional hyperdiffusion and hyperBurgers model as shown by \citet{Boyd94}, who resembles the ``infinite shock'', as the asymptotic Galerkin truncation from hyperviscosity of spectral viscosity, and the ``Gibbs shock'' (the asymptotic approximation to the truncated Fourier series of a discontinuous function, the `Gibbs phenomenon'). Note that it defies analytical solution to the Burgers models to introduce Galerkin truncation and hyperviscosity or other general spectral viscosity \citep{Tadmor89}. Indeed, Galerkin truncated Burgers has intrinsic stochasticity which will approach the absolute equilibrium in the end \citep{AKMCoPAM03}, and the residue of thermalization leads to the bottleneck phenomena in the hyperviscous simulation, like that of Navier-Stokes, in between the inertial and dissipation ranges \citep{FrischPRL08}. \citet{SagautGermano05} observed that filtering a shock introduces parasitic contributions which can seriously spoil the small-scale turbulence.
Nevertheless, various fingerprints of thermalization may be found in turbulence \citep[][and references therein]{hydrochirality}, suggesting the potential of helical absolute equilibrium for understanding shock dynamics in the possibly helical three-dimensional Burgers' turbulence ($c\to 0$), generalized for rotational $\bm{u}$ (which in the traditional definition is the gradient of a potential), or compressible turbulence with pressure.

\section{Effects of polarized thermalization and dissipation in turbulence}

The dissipation rate for the energy of the transverse-velocity ($\hat{u}_{\pm}$), longitudinal-velocity  ($\hat{u}_{|}$) and density ($\zeta$) modes are different, respectively $\eta k^2$, $\frac{4}{3}\eta k^2$ and $0$ (neglecting the `second'/`bulk' viscosity). Besides the small-excitation assumption, K55 further assumes the modes ``exchange energy very rapidly'' and concludes that the `average' dissipation rate of `turbulent [solenoidal/vorticity-mode] energy', $\eta k^2$, be larger than that of `acoustic energy', $\frac{2}{3}\eta k^2$, thus slower decay of turbulence with `noise' than that without `noise'. The theoretical assumption concerning time scales should better be related to the spacial scales/wavelengths (see below).
Note however that the smaller `average' dissipation rate of acoustic modes is due to zero dissipation of the density modes, and the compressive velocity modes themselves have larger dissipation rate than that of the vorticity modes: $\frac{4}{3}\eta k^2>\eta k^2$.
It is interesting to examine the simulation results by \citet{WangETC13}, though it is not completely clear how much relevant are the adiabatic and small-excitation assumptions made in the calculation: In their simulations the total compressive-mode energy is much larger than the vortical-mode one, which however happens in the largest scales and which might be due to the forcing mechanism (they forced on both modes but did not tell further details which can also enter to shape the spectra); while in the inertial range with well-developed turbulence it is shown that the compressive-mode energy level is lower than that of the vortical modes, with apparently steeper scaling law. We now try to understand these observations by the polarized thermalization and the polarized dissipation. The time scales between the quadratic and linear terms in Eq. (\ref{eq:z1}) and (\ref{eq:z2}) balanced (`critical balance' assumption as in magnetohydrodynamic turbulence \citep{GS95} as our tentative step for theoretical interpretation), the eddy turnover time and/or the time scale of acoustic-mode interaction/conversion can be estimated to be $1/(ck)$, while the dissipative time scale for the compressive mode $\hat{u}_|$ is $(\frac{4}{3}\eta k^{2})^{-1}$. When the dissipative time scale enters and
\begin{equation}\label{eq:t1}
1/(ck)<(\frac{4}{3}\eta k^{2})^{-1},
\end{equation}
the internal exchange between the density and compressive modes is faster than dissipation, thus K55's `average'-dissipation-rate argument works, with slower decay of the acoustic modes as appears in \citet{WangETC13}'s simulation with normal viscosity; while in Wang et al.'s simulation with hyperviscosity, with the exponent $n=8$ in $k^{2n}$, the dissipation quickly becomes faster than the internal conversion with
\begin{equation}\label{eq:t2}
1/(ck)>(\frac{4}{3}\eta k^{2n})^{-1}
\end{equation}
as $k$ increases, thus K55's `average'-dissipation-rate argument won't work, resulting in faster decay of the energy spectrum of the compressive mode than the solenoidal mode. If the flow is strongly helical, then polarized thermalization effect may enter: in the former case with normal viscosity, vortical mode should be stronger with the decay slowed down, competing with polarized dissipation effect; in the latter case with hyperviscosity and when K55's argument does not work, both the polarized dissipation and the polarized thermalization favor the vortical mode, leading to the strengthened departure between the vortical and dilatational mode. Thus, the speculation by \citet{WangETC13} that ``As Mach number increases, the interactions between the solenoidal and compressive modes will become stronger, probably leading to a tendency of energy equipartition
between the two modes'', suggested by K55 as the authors explicitly referred to, may not be as favorable for strongly helical compressible turbulence: our Eq. ({\ref{eq:neq}), $U^{\perp} > U^{\sim}$, in the absolute equilibrium state might be extrapolated to more general cases, which could be preliminarily seen by perturbing the last term of Eq. (\ref{eq:ce}) and simply check the corresponding results. Unfortunately, it seems to us that no documentation of compressible helical turbulence data has been made available, and relevant in-silico or laboratory experiments are called for.

\citet{BertoglioBatailleMarionPoF01} reported in the two-point closure theory (with also the extra small-excitation assumption, represented in the low Mach number and the neglection of density fluctuations or entropy modes etc., in their ``EDQNM'' approach) and simulations, forced only on the solenoidal modes, asymptotic equipartition between the kinetic energy and the potential energy of the acoustic (`compressible') mode (Fig. 17 there), which is consistent with the absolute equilibrium equipartition, suggesting that K55's assumption that the acoustic modes `exchange energy very rapidly' may work in their model and simulation so that the dissipative polarization is not effective. However, the EDQNM closure involves approximations or \textit{ad hoc} assumptions, especially as proposed in the subsequent study by Fauchet and Bertoglio [according to the detailed discussions by \citet{SagautCambon08}], more careful treatment for the renormalization of the dispersion frequency of acoustic waves may be necessary and can lead to the breakdown of the so-called ``strong acoustic equilibrium'', the equipartition between the dilatation and density modes at each $k$ as appears in our statistical absolute equilibrium spectra and in the purely linear-wave regime \citep{SagautCambon08}. Absolute equilibrium concerns the ultimate state of thermalization whose times scale is not necessarily much smaller than the turbulence eddy-turn-over time scale, so it would not be surprising that such precise ``strong acoustic equilibrium'' break down in the turbulent state as a result of delicate balance among various mechanisms with different time scales. If we relate the turbulence eddy turn-over time to the de-correlation time modeled by the eddy-damping de-correlation function, then indeed the traditional exponential one by \citet{BertoglioBatailleMarionPoF01} is much slower than the Gaussian one by Fauchet and Bertoglio \citep{SagautCambon08}, so the model of \citet{BertoglioBatailleMarionPoF01} may allow enough thermalization time to have the strong acoustic equilibrium. Note that we have related the strong acoustic equilibrium state to the statistical equilibria, besides the classical linear waves \citep{SagautCambon08}.
So, our results could provide insights for practical use, such as for such two-point closure models or the sub-grid-scale ones \citep{SagautGermano05} which might either amplify or depress the thermalization effects. In the helical case, for instance, artificial treatments (smoothing, hyperviscosity etc.) may strengthen the vortical mode and possibly unphysically (larger) conversion of acoustic energy to vortical energy, leading to underestimating the noise level. The techniques, such as the Kraichnan transformation and the helical representation in the sharpened Helmholtz decomposition, in the treatment here may be applied to further the extensions of \citet{Lee1952} by \citet{FrischETC75}, \citet{ServidioETC08} and \citet{hydrochirality} to compressible magnetohydrodynamics, for similar discussions of the isotropic polarization issue etc.

Concerning fundamental statistical physics, Eq. (\ref{eq:de}) provides some basis and motivates thoughts for the possibility of understanding in a more quantitatively definite way the polarization discussed here, by extending the well-known She-Jackson and Gallavotti-Cohen approaches \citep[][and references therein]{nonir}. The idea is to replace the damping parameters in different components of the Navier-Stokes equations with some time-dependent `dynamical eddy-viscosity' coefficients (such that the system is also time-reversible, but, unlike the Galerkin-truncated invisid system, does not satisfy the Liouville theorem), to respect more observable(s), such as the helicity contained in subsets, say, each shell (She-Jackson), or even the whole set of $\bm{k}$ (Gallavotti-Cohen), besides the energy. The Gallavotti-Cohen approach needs further theoretical assumptions, such as the `chaotic hypothesis' and `equivalent-ensemble conjecture' \citep{nonir}: Such work for the simpler case of incompressible flow is formally easier to handle and is underway. Like the absolute equilibrium calculation, it may not be impossible that quantitatively different fluctuation relations for polarized components of the field could be established to definitely measure the compressible turbulence (isotropic) polarization.

%
%
%

\section*{Acknowledgments}
The author acknowledges communications with C. Cambon on the issues of polarization in anisotropic turbulence, which was partially the motivation of this work, and with P. Sagaut on compressible EDQNM. The discussions on the theoretical possibility of helical non-equilibrium and irreversible dynamical ensembles of time-reversible systems, in response to an anonymous referee's report, have benefited from interactions with L. Biferale and G. Gallavotti.


\bibliographystyle{jfm}


\end{document}